\documentclass[referee]{raa}
\usepackage{graphicx,times}
\usepackage{natbib}
\usepackage{amssymb,amsmath}
\bibpunct{(}{)}{;}{a}{}{,}

\usepackage[a4paper=true,dvipdfm=true,pagebackref=true]{hyperref}
\hypersetup{pdftitle = The title of my PDF, pdfauthor = My name, pdfsubject= The subject, pdfkeywords = keyword1 keyword2 keyword3}
\hypersetup{colorlinks = true, linkcolor = green, anchorcolor = red, citecolor = blue, filecolor = red, pagecolor = red, urlcolor = red}

\begin{document}

   \title{Multi-wavelength emission from 3C 66A: clues on its redshift and gamma-ray emission location
$^*$
\footnotetext{\small $*$ Supported by the National Natural Science Foundation of China.}
}

 \volnopage{ {\bf 2012} Vol.\ {\bf X} No. {\bf XX}, 000--000}
   \setcounter{page}{1}

   \author{Da-Hai Yan\inst{1}, Zhong-Hui Fan\inst{1}, Yao Zhou\inst{1}, Ben-Zhong Dai
      \inst{1}$^\dagger$
\footnotetext{\small $\dagger$ Corresponding author.}
   }

      \institute{ Department of Physics, Yunnan University, Kunming 650091,
China; {\it yandahai555@gmail.com}; {\it fanzh@ynu.edu.cn}; {\it bzhdai@ynu.edu.cn}\\
\vs \no
   {\small Received xxxx; accepted xxxx}
}

\abstract{The quasi-simultaneous multi-wavelength emission of TeV blazar 3C 66A is
studied by using a one-zone multi-component leptonic jet model. It is found that the
quasi-simultaneous spectral energy distribution (SED) of 3C 66A can be well reproduced, especially its {\it Fermi}-LAT first 3 months
average spectrum can be well reproduced by the synchrotron-self Compton (SSC) component plus
external Compton (EC) component of the broad line region (BLR).
Clues on its redshift and gamma-ray emission location are obtained.
The results indicate the following. (i) On the redshift; The theoretical intrinsic TeV spectra can be
predicted by extrapolating the reproduced GeV spectra. Through comparing this extrapolated TeV
spectra with the extragalactic background light (EBL) corrected observed TeV spectra, it is suggested that the
redshift of 3C 66A could be between 0.1 and 0.3, the most likely value is $\sim$ 0.2. (ii) On the gamma-ray emission location; To well reproduce the
GeV emission of 3C 66A under different assumptions on BLR,
the gamma-ray emission region is always required to
be beyond the inner zone of BLR. The BLR absorption effect on gamma-ray emission confirms this point.
\keywords{BL Lacertae objects: individual (3C 66A) --- galaxies: active --- gamma
rays: theory --- radiation mechanisms: non-thermal}
}

   \authorrunning{D.-H. Yan et al. }            
   \titlerunning{Clues on redshift and gamma-ray emission location of 3C 66A}  
   \maketitle

%
\section{Introduction}           
\label{sect:intro}

Blazars are the most extreme class of active galactic nuclei (AGN).
Their SEDs are characterized by two
distinct bumps. The low-energy component originates in relativistic electron synchrotron
emission. The high-energy component could be produced by inverse
Compton (IC) scattering \citep{bott07}.
Various soft photon sources seed SSC
process \citep[e.g.,][]{rees, Maraschi} and external
Compton (EC) process \citep[e.g.,][]{dermer93, Sikora94} in the jet to produce $\gamma$-rays. Hadronic models
have also been proposed to explain the multi-band emissions of
blazars \citep[e.g.,][]{Mannheim, Mucke}.

TeV photons emitted by blazars are
absorbed through the pair-production process, by interaction with
EBL \citep{Stecker92}. The absorption effect depends on both the EBL photon density and the redshift
of the TeV source. The energy range of interest for background
photons here is from optical to ultraviolet (UV). Since it is difficult to
measure the EBL directly, many EBL models are proposed: such as low limit models \citep[e.g.,][]{kneis,raze}, mean
level ones  \citep[e.g.,][]{finke10,Franceschini}, and high level ones
\citep[e.g.,][]{stecker}. \citet{aharon} discussed some gamma-ray
blazars with unexpectedly hard spectra at relative large redshift, and
suggested that EBL is of the first type. \citet{albert} found that
the universe is more transparent to gamma-rays. However,
\citet{stecker09} pointed out that \citet{albert} do not
significantly constrain the intergalactic low energy photon spectra
and their high level EBL model is still valid. In an analysis of
photons above 10 GeV from gamma-ray sources detected by
\textit{Fermi}-LAT, \citet{abdoa} found evidence to exclude the high
level EBL models. The EBL absorption effect on gamma-rays is helpful to constrain
the redshift of TeV sources.
For instance, the SED of a blazar can be extrapolated into the TeV region
by reproducing the
multi-band (optical-GeV band) data with certain emission model. The redshift
of the VHE source can then be constrained by comparing
the EBL-corrected observed TeV spectrum with the extrapolated one.

It's well known that the high energy emissions of some blazars need EC components.
The energy density of
external photon field is related to the gamma-ray emission location \citep[e.g.,][]{g09}.
Therefore, the clue on the gamma-ray emission region location of a blazar can be obtained
from its high energy emission \citep[e.g.,][]{yan}. Moreover,
the external photons absorption on the gamma-ray emission is also helpful to constrain the gamma-ray emission location of blazar \citep[e.g.,][]{liu,bai,pout}.

3C 66A is classified as intermediate BL Lac (IBL), because of its
synchrotron peaking between $10^{14}$ Hz and $10^{15}$ Hz
\citep{perri,abdob}. The most widely used redshift for 3C 66A is
0.444, based on a single emission line measurement \citep{miller}.
However, \citet{miller} stated that they were not sure of the
reality of this emission feature, and warned that the redshift is
not reliable. Later, \citet{lanze} confirmed the redshift of 0.444
based on data from \textit{International Ultraviolet Explorer}
(\textit{IUE}). However, \citet{bramel} argued that the 3C 66A
redshift determined using \textit{IUE} data is questionable.
\citet{finke2} placed a lower limit on the redshift of 3C 66A,
$z\geq0.096$, using information regarding its host galaxy. Recently,
\citet{pran} suggested that the redshift of 3C 66A should be below
$0.34\pm 0.05$, and that the most likely redshift is $0.21\pm0.05$, by assuming that the EBL-corrected TeV spectrum are not
harder than the \textit{Fermi}-LAT spectrum.

\citet{Joshi} suggested that $\gamma$-ray emission of 3C 66A in the flare state could be dominated by
an external Compton (EC) process. \citet{yang} found that the TeV
emission has contribution from EC when taking $z=0.444$, or by pure SSC
when $z=0.1$. \citet{abdo11} studied the SED of 3C
66A at flare state by using the SSC+EC model, and suggested that the
redshift of 3C 66A may be between 0.2 and 0.3.

A quasi-simultaneous multi-wavelength observations campaign for
3C 66A was carried out by \textit{Fermi} and \textit{Swift} from
August 2008 to October 2008. VERITAS observed 3C 66A for 14 hours from 2007 September
through 2008 January and for 46 hours between 2008 September and
2008 November \citep{accia3c,acciaa}. In this work, The {\it Fermi}-LAT first 3 months
average spectrum and the VERITAS average spectrum based on the observations from 2007 September through
2008 November
are used. Data from the radio, optical, UV,
X-ray, and GeV $\gamma$-ray to TeV $\gamma$-ray bands are publicly available
\citep{abdob}. In this work, we study the
quasi-simultaneous SED of 3C 66A with a multi-component
leptonic jet model, and constrain its redshift and gamma-ray emission location.
We adopt the cosmological parameters ($H_0, \Omega_m,
\Omega_{\Lambda}$) = (70 km s$^{-1}$ Mpc$^{-1}$, 0.3, 0.7)
throughout this paper.


\section{The model}
\label{sect:model}

We assume that multi-band emission of a blazar is produced in a spherical blob
in the jet, which is moving relativistically at a small angle to our
line of sight. The observed radiation is strongly boosted by a
relativistic Doppler factor $\delta_D$. The relativistic electrons
inside the blob lose energy via synchrotron emission and IC
scattering. The electron distribution is \citep{dermer09},
\begin{eqnarray}
N_{\rm e}^{\prime}(\gamma^{\prime})=K_{\rm
e}^{\prime}H(\gamma^{\prime};\gamma_{\rm min}^{\prime},\gamma_{\rm
max}^{\prime}){\gamma^{\prime -p_1}exp(-\gamma^{\prime}/\gamma_{\rm
b}^{\prime})}
\nonumber \\
\times H[(p_{\rm 2}-p_{\rm 1})\gamma_{\rm
b}^{\prime}-\gamma^{\prime}]+[(p_{\rm 2}-p_{\rm 1})\gamma_{\rm
b}^{\prime}]^{p_{\rm 2}-p_{\rm 1}}\gamma^{\prime -p_{\rm 2}}
\nonumber \\
\times exp(p_{\rm 1}-p_{\rm 2})H[\gamma^{\prime}-(p_{\rm 2}-p_{\rm
1})\gamma_{\rm b}^{\prime}]K_{\rm
e}^{\prime}H(\gamma^{\prime};\gamma_{\rm min}^{\prime},\gamma_{\rm
max}^{\prime}),
\end{eqnarray}
where $K^{\prime}_{\rm e}$ is the normalization factor, which describes the
number of relativistic electrons in emitting blob.
$H(x;x_{1},x_{2})$ is the Heaviside function:
$H(x;x_{1},x_{2})=1$ for $x_{1}\leq x\leq x_{2}$ and
$H(x;x_{1},x_{2})=0$ everywhere else; as well as $H(x)=0$ for $x<0$
and $H(x)=1$ for $x\geq0$. In the co-moving frame, this
distribution is a double power law with two energy cutoffs:
$\gamma_{\rm min}^{\prime}$ and $\gamma_{\rm max}^{\prime}$. The
spectrum is smoothly connected with indices $ p_1$ and $p_2$ below
and above the electrons' break energy $\gamma_{\rm b}^{\prime}$.
Note that here and throughout the paper, unprimed quantities
refer to the observer's frame and primed ones refer to the co-moving
frame.

The multi-component model of \citet{dermer09} is used to reproduce
the SED of 3C 66A. For EC components,
we consider photons emitted directly from the accretion disk
and photons from the central source Thomson scattered at BLR as the seed photons.
In addition, we take into account gamma-ray
attenuation by the BLR-scattered radiation field.

We assume that the BLR is a spherically symmetric shell with inner
radius $R_i$ and outer radius $R_{\rm o}$. It's assumed that the
gas density of the BLR has the power-law distribution $n_{\rm
e}(r)=n_0(\frac{r}{R_{\rm i}})^\zeta$, where $ R_{\rm i}\leq r \leq
R_{\rm o}$. The radial Thomson depth is given by $\tau_{\rm
T}=\sigma_{\rm T}\int^{R_{\rm o}}_{R_{\rm i}}dr n_{\rm e}(r)$, where
$r$ is the distance from the central black hole \citep{dermer09}. In
our calculation, we use $\tau_{\rm T}=0.01$, which is the typical
value \citep{finke102,reimer,donea}. \citet{kaspi}
suggested that the particle density of BLR scales as $r^{-1.0}$ or
$r^{-1.5}$. In our calculation, we adopt the exponent $\zeta=-1.0$.

Using reverberation mapping, \citet{bentz} derived an improved
empirical relationship between BLR radius $R_{\rm BLR}$ and
luminosity $L_{\lambda}$ at $5100 {\rm \AA}$:
\begin{equation}
{\rm log}(R_{\rm BLR})=-21.3+0.519\cdot {\rm log}(\lambda
L_{\lambda}(5100 {\rm \AA}))\,.
\end{equation}
The V-band magnitude of 3C 66A is 15.21 \citep{veron}. We use the
optical spectral index given by \citet{fiorucci} to calculate the
average flux at $5100 {\rm \AA}$, which is 2.785 mJ. In this work,
we take the estimated $R_{\rm BLR}$ as the outer radius of the BLR
$R_{\rm o}$. \citet{peter} suggested that the typical size of the
BLR in quasars is on the order of light-months. We follow several
authors \citep{reimer,donea}, using the relationship
$R_{\rm i}=R_{\rm o}/40$ to derive a value for $R_{\rm i}$.

\begin{figure}
   \centering
   \includegraphics[width=14.0cm, angle=0]{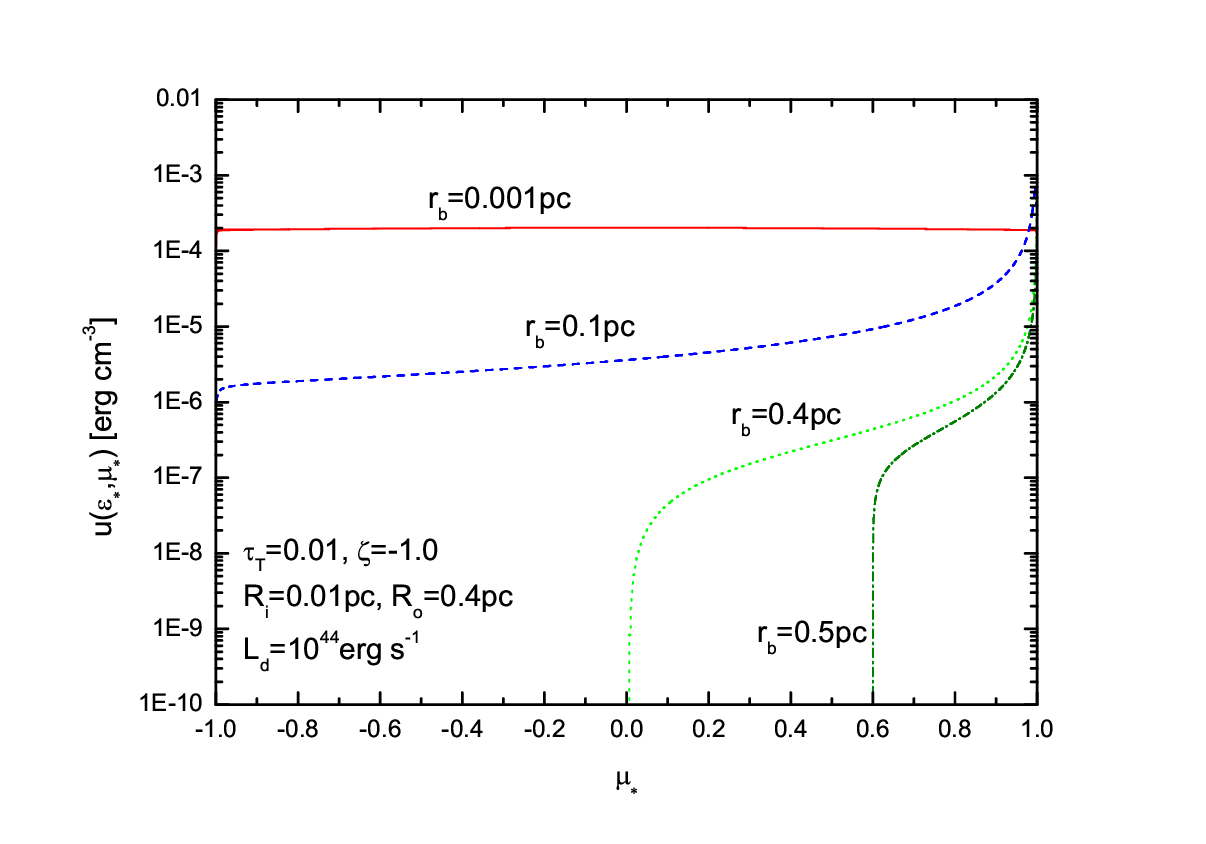}
   \caption{Angle-dependent energy density of BLR-scattered photon
field. The values of $r_{\rm b}$ are labeled on the curves. The dimensionless photon energy is $\epsilon_{\ast}=2.48\times10^{-5}$. }
   \label{Fig1}
   \end{figure}

To simplify calculation, the BLR-scattered photon field is assumed
to be monochromatic with energy $\epsilon_{\ast}$, which is the mean
energy from the accretion disk \citep{dermer09}. The approximation for the
mean dimensionless photon energy from a standard
accretion disk \citep{ss} at radius $R$ is given by \citep[e.g.,][]{dermer09,finke102}
\begin{equation}
\epsilon_{\rm d}(R)=1.5\times10^{-4}(\frac{10\ell_{\rm Edd}}
{M_8\eta})^{\frac{1}{4}}(\frac{R}{r_{\rm g}})^{-\frac{3}{4}}\,.
\end{equation}
The accretion luminosity is $\ell_{\rm Edd}=\frac{L_{\rm d}}{L_{\rm
Edd}}$, which here has the value 0.03. The Eddington luminosity is
$L_{\rm Edd}=1.26\times10^{46}M_8 \rm \,ergs\cdot s^{-1}$, and
$L_{\rm d}$ is the accretion disk luminosity. The accretion
efficiency $\eta$ is 0.1. The gravitational radius $r_{\rm
g}=\frac{GM}{c^2}\cong1.5\times10^{13}M_8 \rm \,cm$, where $c$ is the speed of light.
The black hole mass of 3C 66A is $M_8=\frac{M_{\rm BH}}{10^8M_{\rm \odot}}=4.0$
\citep{ghisellini}.
In this work, we adopt $\epsilon_{\ast}=\epsilon_{\rm d}(10r_{\rm
g})=2.48\times10^{-5}$, corresponding to the energy of 13 eV, which is the typical
energy of photons from a standard accretion disk. The energy density of BLR-scattered photon
field is
\begin{equation}
u(\epsilon_{\ast},\mu_{\ast};r_{\rm b})=\frac{L_{\rm d}r_{\rm
e}^2}{3cr_{\rm b}}F(\mu_{\ast},r_{\rm b})
\end{equation}
\citep{dermer09}, where $r_{\rm e}$ is classic electron radius. $r_{\rm b}$ is the distance from the
emission blob to the central black hole. $F(\mu_{\ast},r_{\rm
b})$ is the function given by \citet{dermer09} (their Eq.(97)), which is related to the gas energy density in BLR $n_{\rm e}(r_{\rm b})$. Here, $\tau_{\rm T}$ is used to normalize $n_{\rm e}(r_{\rm b})$. The
energy density of BLR-scattered photon
field is angle-dependent. $\theta_{\ast}$ is the angle between the directions of the BLR scattered photon and motion of blob, which is also the interaction angle between the relativistic electron and soft photon \citep{dermer09}. $\mu_{\ast}$ is the value of cos$\theta_{\ast}$. In Fig.~\ref{Fig1}, we show the energy density of BLR-scattered photon
field, varying with $r_{\rm b}$.

The intrinsic high energy photons flux from extragalactic sources is
\begin{equation}
f_{\rm intrinsic}(E_\gamma)=e^{\tau(E_\gamma,z)}f_{\rm
observed}(E_\gamma) \,,
\end{equation}
where $f_{\rm observed}$ is the measured TeV flux, and
$\tau(E_\gamma,z)$ is the optical depth of $\gamma$-ray with
energy $E_{\gamma}$ at redshift $z$.
Here, we use the EBL model of
\citet{Franceschini}\footnote{Opacities for photon-photon
interaction as a function of the source redshift are available on
the the website http://www.astro.unipd.it/background.} to de-absorb
the observed TeV spectra. This model is based on observations and takes
into account all available information on cosmic sources
contributing background photons.

Several parameters in our model can be constrained by observations.
\citet{bott09} excluded extreme values
of the Doppler factor in the range $\delta_{\rm D} \geq 50$. The
size of the emission blob can be constrained by the observed
variability timescales $t_{\rm var}$, because $R^{\prime}_{\rm b}=t_{\rm
v,min}\delta_D c/(1+z)\leq\delta_D c t_{\rm tar}/(1+z)$. Here
$R^{\prime}_{\rm b}$ is the radius of the blob in the co-moving frame, and
$t_{\rm v,min}$ is the smallest variability timescale.
\citet{takalo} reported a micro-variability with $t_{\rm var}\sim
2.16\times10^4$s and $\bigtriangleup \rm mag\sim0.2$.
\citet{abdo11} reported shorter variability at optical band:
$t_{\rm var}\sim 1.44\times10^4$s.

\section{The results}

\begin{figure}
   \centering
   \includegraphics[width=14.0cm, angle=0]{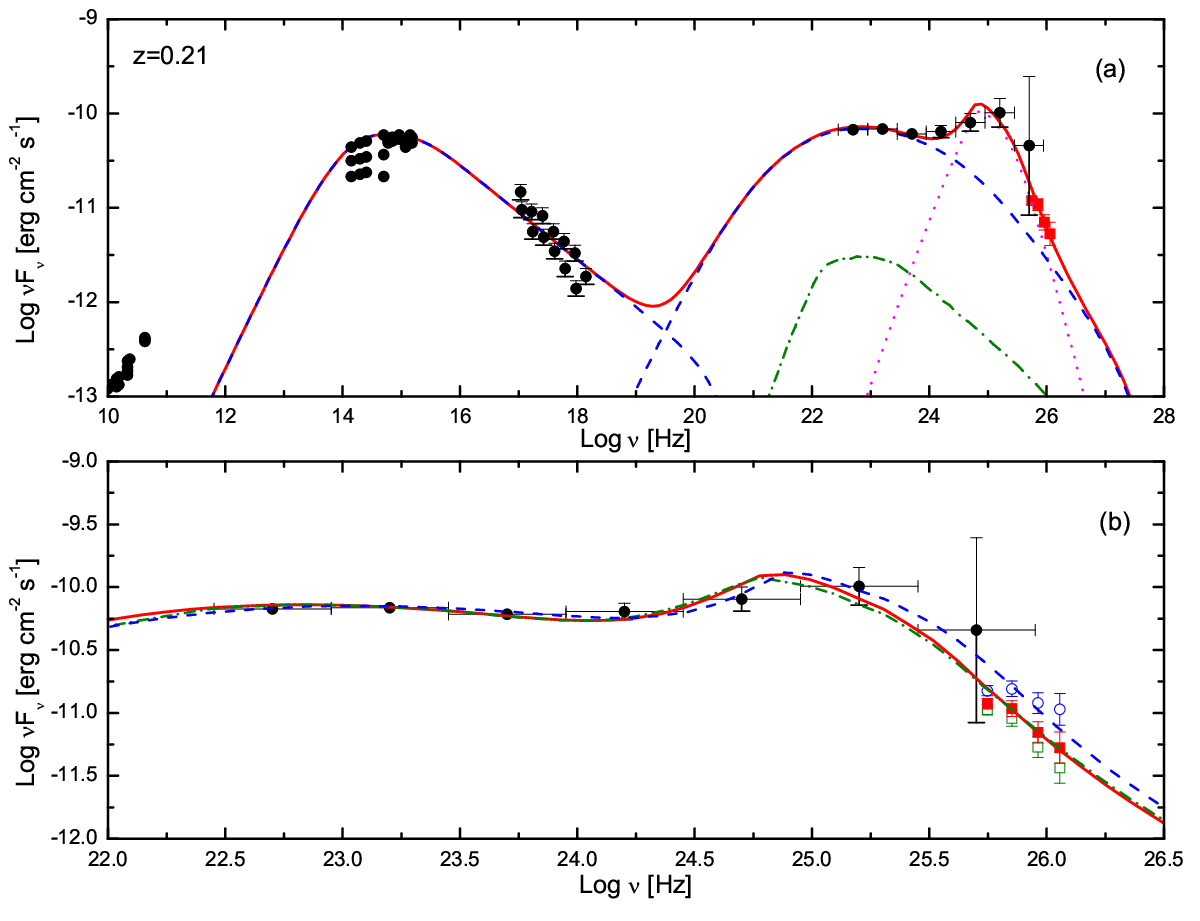}
   \caption{In panel (a), we show the reproduced SED with
$z=0.21$. The filled square are the de-absorbed TeV data with
$z=0.21$. The dashed, dash-doted, dotted and thick solid lines are
SSC component, accretion-disk, BLR-reproduced component and the sum
of multi-component, respectively. In panel (b), the open square,
filled square and open circle are the de-absorbed TeV data with
$z=0.15$, 0.21 and 0.31, respectively. The dash-dotted, solid and
dashed lines are the model results at $z=0.15$, 0.21 and 0.31,
respectively. All observed data are from \citet{abdob}. See detailed data information in \citet{abdob}. }
   \label{Fig2}
   \end{figure}

   \begin{figure}
   \centering
   \includegraphics[width=14.0cm, angle=0]{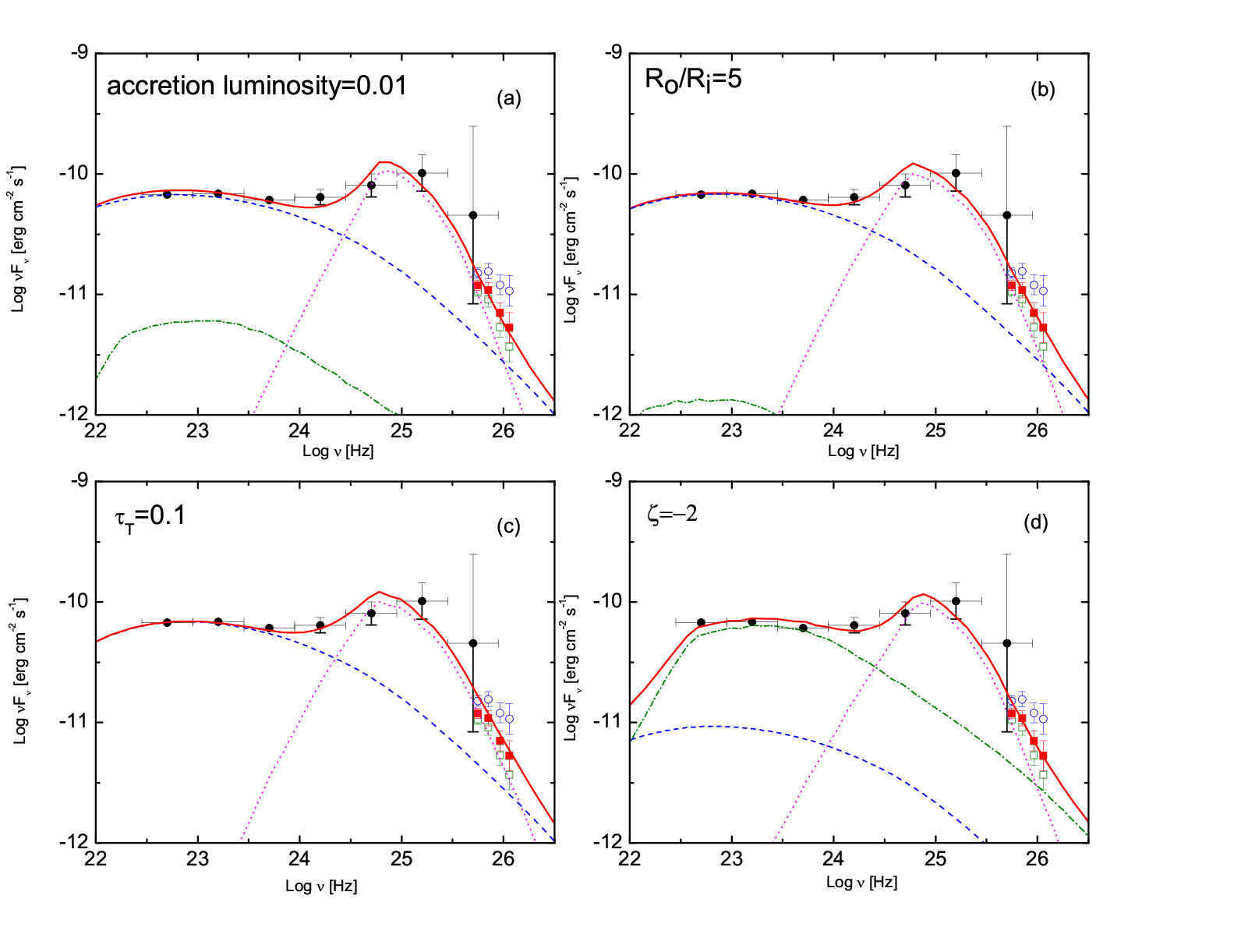}
   \caption{The effects of different assumptions of BLR structure and
the characteristics of central source on the estimation of the
redshift. The symbols are the same as that in Fig. ~\ref{Fig2}. }
   \label{Fig3}
   \end{figure}

   \begin{figure}
   \centering
   \includegraphics[width=14.0cm, angle=0]{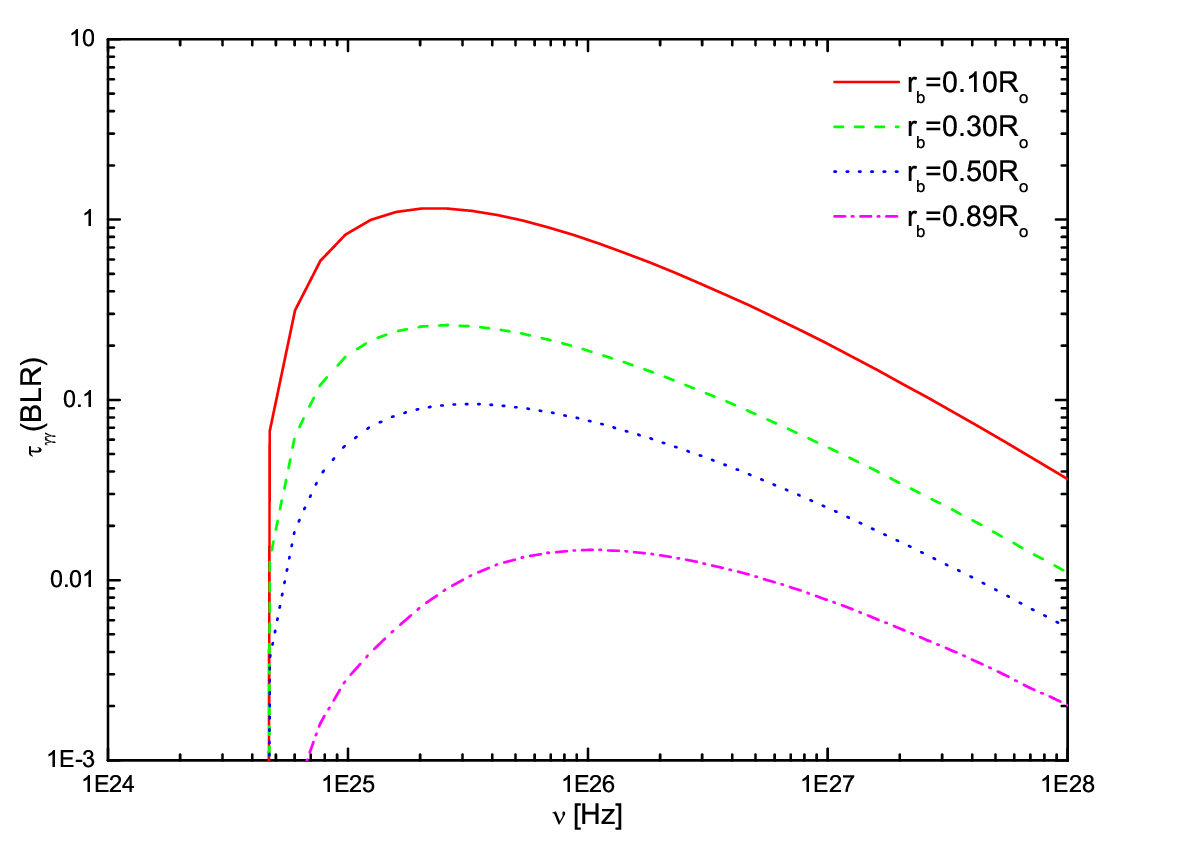}
   \caption{$\gamma \gamma$ optical depth for $\gamma$-ray interaction
with BLR-reproduced photons at different distances from central BH
when $z=0.21$. }
   \label{Fig4}
   \end{figure}

   \begin{table}
\bc
\begin{minipage}[]{100mm}
\caption[]{Model parameters for Fig.~\ref{Fig2}. \label{tab}}\end{minipage}
\setlength{\tabcolsep}{2.5pt}
\small
 \begin{tabular}{ccccccccccc}
  \hline\noalign{\smallskip}
parameters & $z=0.15$ & $z=0.21$ & $z=0.31$ \\
 \hline
 B (G) & 0.168 & 0.168 & 0.168\\
 $K^{\prime}_{\rm e}$ ($10^{53}$) & 0.62 & 1.5 & 1.5\\
 $p_1$ & 2.0 & 2.0 & 2.0\\
 $p_2$ & 4.0 & 4.0 & 4.0\\
 $\gamma_{\rm max}^{\prime}$($10^6$)& 3.0 & 3.0 & 3.0\\
$\gamma_{\rm b}^{\prime}$($10^3$) & 5.8 & 6.3 & 7.6\\
$\gamma_{\rm min}^{\prime}$($10^3$) & 1.93 & 1.90 & 1.76\\
$\delta_D$ & 38 & 36 & 43\\
$t_{\rm v, min}$($10^4$s) & 0.69 & 1.17 & 1.21\\
\hline
$M_{\rm 8}$ & 4.0 & 4.0 & 4.0\\
$\ell_{\rm Edd}$ & 0.03 & 0.03 & 0.03\\
$\eta$ & 0.1 & 0.1 & 0.1\\
 \hline
 $\tau_{\rm T}$ & 0.01 & 0.01 & 0.01\\
 $\zeta$ & -1.0 & -1.0 & -1.0\\
 $R_{\rm i}$ ($10^{-2}$pc)& 0.25 & 0.35 & 0.55\\
 $R_{\rm o}$ (pc)& 0.1 & 0.14 & 0.22\\
 $r_{\rm b}$ ($R_{\rm o}$)& 1.03  & 0.89 & 0.72\\
  \noalign{\smallskip}\hline
\end{tabular}
\ec
\end{table}

\begin{table}
\bc
\begin{minipage}[]{100mm}
\caption[]{Model parameters for Fig.~\ref{Fig3}. \label{tab1}}\end{minipage}
\setlength{\tabcolsep}{2.5pt}
\small
 \begin{tabular}{ccccccccccc}
  \hline\noalign{\smallskip}
parameters & $\ell_{\rm Edd}=0.01$ & $\frac{R_{\rm o}}{R_{\i}}=5$ & $\tau_{T}=0.1$ & $\zeta=-2$\\
 \hline
 B (G) & 0.168 & 0.168 & 0.168 & 0.168\\
 $K^{\prime}_{\rm e}$ ($10^{53}$) & 1.5 & 1.5 & 1.6 & 1.5\\
 $p_1$ & 2.0 & 2.0 & 2.0 & 2.0\\
 $p_2$ & 4.0 & 4.0 & 4.0 & 4.0\\
 $\gamma_{\rm max}^{\prime}$($10^6$)& 3.0 & 3.0 & 3.0 & 3.0\\
$\gamma_{\rm b}^{\prime}$($10^3$) & 6.3 & 6.3 & 5.6 & 6.3\\
$\gamma_{\rm min}^{\prime}$($10^3$) & 1.8 & 2.0 & 2.5 & 1.9\\
$\delta_D$ &36  & 36 & 37 & 36\\
$t_{\rm v, min}$($10^4$s) & 1.2 & 1.17 & 1.05 & 3.2\\
\hline
$M_{\rm 8}$ & 4.0 & 4.0 & 4.0 &4.0\\
$\ell_{\rm Edd}$ & & 0.03 & 0.03 & 0.03\\
$\eta$ & 0.1 & 0.1 & 0.1 & 0.1\\
 \hline
 $\tau_{\rm T}$ & 0.01 & 0.01 & & 0.01\\
 $\zeta$ & -1.0 & -1.0 & -1.0 & \\
 $R_{\rm i}$ ($10^{-2}$pc)& 0.35 & 2.8 & 0.35 & 0.35\\
 $R_{\rm o}$ (pc)& 0.14 & 0.14 & 0.14 & 0.14\\
 $r_{\rm b}$ ($R_{\rm o}$)& 0.65  & 1.02 & 1.31 & 0.52\\
  \noalign{\smallskip}\hline
\end{tabular}
\ec
\end{table}

In Fig. \ref{Fig2}, we show the modeling results at three different redshifts.
The filled circles are quasi-simultaneous data from radio to GeV. The observed VERITAS data are
EBL-corrected by using the EBL model of \citet{Franceschini} with
different redshifts. It can be seen that the accretion-disk component is negligible compared to
the SSC and BLR components. SSC and EC are responsible for emissions
at the GeV-TeV bands. Emission between 0.1 GeV and 10 GeV is dominated
by SSC. Above 10 GeV, the EC component of BLR is more important.
Table~\ref{tab} lists all model parameters.

It is interesting that the Klein-Nishina (KN) effect
becomes important in Compton scattering
the BLR radiation when $\gamma^{\prime}\Gamma_{\rm bulk}\epsilon_{\ast}\geq1/4$,
where $\Gamma_{\rm bulk}$ is the bulk Lorentz factor of the blob. In our model,
$\Gamma_{\rm bulk}\approx\delta_{\rm D}$, so that $\gamma^{\prime}_{\rm KN}=280$.
Electrons with this energy scatter photons
primarily to energies of $\epsilon_{\rm KN}\approx\Gamma_{\rm bulk}\delta_{\rm D}\epsilon_{\ast}\gamma^{\prime 2}_{\rm KN}/(1+z)\approx2.08\times10^3$, which corresponds to frequency of $\nu_{\rm KN}\approx2.57\times10^{23}$Hz.
Due to the KN effect, the BLR-component spectra at the right side of peak decline more quickly.
In addition to large $\gamma_{\rm min}^{\prime}$, the KN effect is the other cause of the narrow BLR-component SED.

As shown in panel (b) of Fig.~\ref{Fig2}, the EBL-corrected TeV
spectrum is steeper than the extrapolated one if the redshift
is below 0.15. On the other hand, if the redshift is above
0.31, the EBL-corrected TeV spectra becomes harder. The EBL-corrected TeV emission can be well reproduced when z=0.21.
Hence, the redshift of 3C 66A should be between 0.15 and 0.31, and the most likely redshift is 0.21.
There are several poorly constrained parameters
in our model. It should be discussed whether the uncertainties of model parameters can
affect our results. As mentioned above,
the contribution of the BLR component is dominant at TeV band, which is crucial for
constraining the redshift of 3C 66A. The BLR structure ($R_{\rm i}$,
$R_{\rm o}$, $\zeta$, $\tau_{\rm T}$) and the characteristics of the
central source (the black hole and its accretion disk) can affect
the contribution of the BLR component. $R_{\rm o}$ can be
constrained by Eq.(2). We assumed typical values: $(\ell_{\rm Edd}, R_{\rm
o}/R_{\rm i}, \tau_{\rm T}, \zeta )= (0.03, 40, 0.01, -1)$, to reproduce the SED of 3C 66A.
The effects of these parameters on estimating of the redshift are discussed by using other plausible boundary values.
Results are shown in Fig. \ref{Fig3} (a), (b), (c) and (d).
Parameters are listed in Table~\ref{tab1}.
For clarity, only the modeling results in the high energy part
of the case $z=0.21$ are shown. Obviously, the SED (including TeV spectra) can also be reproduced well.
We therefore argue that our results are independent of these parameters.

In addition, our results indicate that the gamma-ray emission region is beyond the inner zone of BLR ($\sim$0.1 pc, see Table~\ref{tab}~\ref{tab1}).
In Fig. \ref{Fig4}, we show the $\gamma \gamma$ absorption by
BLR-scattered radiation at different blob locations when
taking $z=0.21$. There is a significant absorption when the blob is
inside the inner zone of the BLR. Beyond the inner zone, absorption
is negligible. No absorption feature at GeV band confirms that the emission region of 3C 66A should be out of the inner zone of BLR.

\section{Discussion and conclusion}
\label{sect:discussion}

A pure SSC model fails to explain the average GeV spectrum of 3C 66A observed
by \textit{Fermi}-LAT during its first three months operation.
While, a satisfactory reproduction of the data can
be obtained by the multi-component model (see Fig. \ref{Fig2}~\ref{Fig3}), which
takes into account not only the specific shell structure
of the BLR, but also the angular dependence of the photon
distribution. The multi-component model requires a large value of $\gamma_{\rm
min}^{\prime}\sim2\times10^3$. As argued by \citet{tavecc09}, this
result seems to provide important clues to the electron
acceleration process and the role of energy loss. A large value of
$\gamma_{\rm min}^{\prime}$ leads to a steep spectrum in the
low-energy band, so our model does not explain the observed radio
emission. The radio emission may come from a larger emission region.

Based on the modeling results, we try to constrain the redshift of 3C 66A through connecting the GeV-TeV spectra.
Because we can not give the error estimate by using this method, we think only the redshift range we derived is significant.
It's therefore suggested that the redshift of 3C 66A may be between 0.1 and 0.3, and the most likely one is $\sim$ 0.2.
Furthermore, we found the results are independent of the assumptions about the BLR structure we made.
By using different emission model and GeV-TeV data, we obtained the very similar results with that obtained by \cite{abdo11}.
However, it should be kept in mind that both our results and that of \cite{abdo11} depend on the EBL model.
We also try to get clues on the gamma-ray emission location of 3C 66A. Combining the BLR absorption effect and the EC component required to reproduce the gamma-ray emission, our results indicate that the gamma-ray emission region of 3C 66A may be in the outer zone of BLR or out of BLR.

\normalem
\begin{acknowledgements}
We thank the referee for the constructive comments.
We thank L. Zhang and X. W. Cao for helpful comments on this paper and J. P. Yang
for helpful discussion. This work is supported by the National
Science Foundation of China (grants 11063003 and 10963004), Yunnan
Provincial Science Foundation (grant 2009CI040).

\end{acknowledgements}

\bibliographystyle{raa}
\bibliography{bibtex}

\end{document}